\documentclass{article}

\usepackage{arxiv}

\usepackage[utf8]{inputenc} 
\usepackage[T1]{fontenc}    
\usepackage{hyperref}       
\usepackage{url}            
\usepackage{booktabs}       
\usepackage{amsfonts}       
\usepackage{nicefrac}       
\usepackage{microtype}      
\usepackage{lipsum}		
\usepackage{graphicx}
\usepackage[square,numbers]{natbib}
\usepackage{doi}

\title{Spatially-resolved coherence of organic molecular spins at room-temperature}

\author{ {Adrian ~Mena$^1$} \\
	\And
    {Nicholas P. Sloane$^1$}
    \And
    {Max R. Bonengel$^1$}
    \And
    {Dane R. McCamey*$^1$}
}

\date{}

\renewcommand{\headeright}{}
\renewcommand{\undertitle}{}
\renewcommand{\shorttitle}{}

\hypersetup{
pdftitle={Spatially-resolved coherence of organic molecular spins at room-temperature},
pdfsubject={q-bio.NC, q-bio.QM},
pdfauthor={David S.~Hippocampus, Elias D.~Striatum},
pdfkeywords={Quantum Sensing, Thin-films, Disorder, Molecular Spins},
}

\begin{document}
\maketitle
\begin{center}
    $^1$ School of Physics, UNSW Sydney, Sydney, 2052, NSW, Australia
    \vspace{1.5cm}
\end{center}
\begin{abstract}
	Molecular spins are a versatile platform for quantum sensing. Not only are the spin-bearing molecules themselves widely tunable, they are also capable of being used as sensors as crystals, films and in solution. Using thin-films offers the advantages of high doping ratios and the ability to control the thickness with nanometre precision, however they also introduce disorder to the system. High proximity sensing can also be realised by using micro- and nano-crystals, however in many solid-state systems this leads to a reduction in coherence. In this paper we combine room-temperature optically-detected coherent control of molecular spins and microscopy to image the coherence properties of both thin-films and micro-crystals of pentacene doped \textit{p}-terphenyl. In thin-films we find large amounts of variation in both the contrast and coherence times, leading to a variability in the magnetic field sensitivity of approximately $7.6$\,\%. Applying the technique to micro-crystals shows much lower sensitivity variability ($1.3\,$\%), and we find no evidence of coherence loss toward the edge of the crystal. Finally we perform optically-detected coherent control on a nano-crystal, showing minimal loss in coherence and contrast compared to the bulk crystal, with a coherence time of $1.09\,\mu$s and a contrast of $25\,$\%.
\end{abstract}

\section{Introduction}\label{sec1}

Optically addressable electronic spins have enabled high precision metrology, with measurements of magnetic fields\cite{wolf2015subpicotesla}, electric fields\cite{dolde2011electric, barson2021nanoscale}, crystallographic strain\cite{doherty2014electronic} and temperature\cite{singh2025high} below what can be achieved classically. Defects in crystalline solids such as nitrogen vacancies (NV)\cite{doherty2013nitrogen} in diamond and silicon carbide vacancies\cite{christle2015isolated, miao2019electrically} have enabled the development of room temperature, as their quantum states support long coherence times and high optical contrasts. This has allowed for the study of a range of phenomena, from biological processes\cite{wu2025fluorescent,zhang2024dynamics}, to imaging current densities across semiconducting devices\cite{tetienne2017quantum, scholten2022imaging, wood20243d}, furthering our understanding of microscopic processes. Alongside improving  sensitivity, quantum sensors have focused on improving spatial resolution for imaging by reducing the distance from sensor to target, for example via the use of shallow-doped diamonds\cite{lovchinsky2016nuclear}. Further developments have enabled the miniaturisation of detectors, pushing our imaging capabilities to smaller scale, for example by utilising NV centres embedded in nanodiamonds to study thermal processes within single cells\cite{kucsko2013nanometre, fujiwara2020real}. Although the three dimensional crystal structure of these materials lead to well protected quantum states, their crystallinity creates a physical barrier preventing these devices from reaching the nanometer scale and their integration into nanoscale structures. Motivated by the desire for higher proximity measurements with smaller probes, alternative materials have gained considerable interest, such as defects in the two-dimensional van der Waals (vdW) material hexagonal boron nitride (hBN)\cite{gottscholl2020initialization, stern2022room}, and one-dimensional vdW materials such as boron-nitride nanotubes\cite{gao2024nanotube}. Following from the same motivation, a growing interest in molecular spins for quantum sensing has developed due to their inherent flexible deployment, providing a potential avenue to quantum sensing with thin-films, crystals, solutions and functionalised spin-bearing probes. 

Spin bearing molecules make up a versatile sensing platform, with a number of spin systems including photo-excited metastable triplets\cite{mena2024room, mann2025chemically}, molecular colour centres with ground state triplets\cite{kopp2024luminescent, kopp2025optically, chowdhury2025room, bayliss2020optically}, coupled multi-spin systems\cite{gorgon2023reversible} and engineered fluorescent proteins\cite{feder2025fluorescent, abrahams2024quantum}. Recently molecular spins have been explored as room-temperature quantum sensors for magnetometry\cite{singh2025room, li2025robust} and thermometry\cite{singh2025high, ishiwata2025molecular}. A key benefit of molecules is that they are inherently nanoscale, enabling their integration into structures such as proteins with a reduced impact on their structure and dynamics\cite{bertran2020light}. These spin systems can be deployed in 3D crystalline structures or as evaporated thin-films, with the later as a potential method for achieving high resolution sensing. Doped crystals offer well protected quantum states, however thin-films are particularly attractive for imaging as they can be grown with nanometer precision\cite{kang2014enhancing} and with a high density of spin-bearing dopant molecules\cite{lubert2018identifying}. The trade-off for this over their crystalline counter-parts comes from the introduction of disorder, which has the potential to impact the uniformity of the film's sensitivity and therefore the imaging capability. In this paper we combine room-temperature optically detected coherent control of molecular spins with wide-field microscopy to characterise the sensitivity of both disordered thin-films and crystalline sensors. This investigation aims to understand the different approaches to deploying molecular spins as quantum sensors and guide efforts towards their use in high-proximity quantum sensing.

\begin{figure}[htb!]
{\includegraphics[width=80mm]{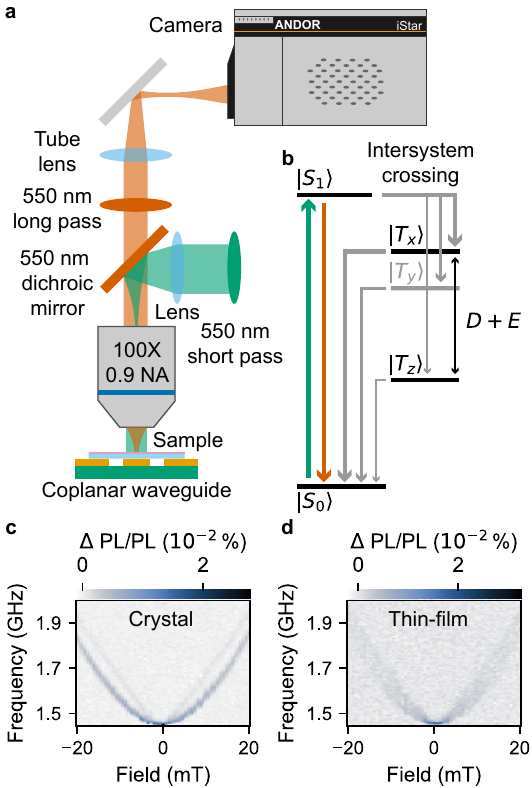}}
\caption{\textbf{Optically detected magnetic resonance with organic triplets.} a) Schematic of the setup used for spatially resolved optically detected magnetic resonance, including a microwave source, optical excitation, emission, optics for imaging and a camera used to perform spatially resolved measurements. b) Energy structure for pentacene, with a singlet ground and excited state. The lowest lying triplet is an excited state which can be populated via intersystem crossing. The triplet states are spin-1 and have microwave addressable transitions. A field-frequency sweep of the $|T_x\rangle \leftrightarrow |T_z\rangle$ transition can be seen in c) and d) for the crystal and film respectively.\label{fig1}}
\end{figure}

We explore this with a staple system when it comes to molecular spins, pentacene doped in \textit{p}-terphenyl. Figure \ref{fig1}a shows the essential elements of our experiments; Our sample is mounted on a co-planar waveguide (CPW) which is used to generate microwave frequency fields. The excitation, by a 520\,nm laser, is focused on the back focal plane of the microscope objective, allowing for a flat illumination across the field of view. We collect the photoluminescence, imaging it onto a sCMOS camera which allows us to perform a spatially-resolved optically detected coherent control measurement. This technique allows for parallel readout of a region of either thin-film or crystal samples, providing a rapid testbed to study the coherence properties across both ordered and disordered sensing substrate. Further details regarding the experimental setup can be found in the methods section.

Figure \ref{fig1}b contains a diagram of the key elements that enable optically detected magnetic resonance (ODMR) in pentacene doped \textit{p}-terphenyl (pc:ptp). Photoexcitation drives the singlet ground state ($|S_0\rangle$) into the singlet excited state ($|S_1\rangle$). Either fluorescence back to $|S_0\rangle$ or intersystem crossing into the triplet manifold can occur from $|S_1\rangle$. The triplets generated from intersystem crossing are spin polarised due to anisotropic intersystem crossing rates, with probabilities $P_x:P_y:P_z = 0.76:0.16:0.08$ for the $|T_x\rangle$, $|T_y\rangle$ and $|T_z\rangle$ triplet sublevels respectively\cite{sloop1981electron}. Likewise the depopulation rates are also anisotropic, with the decay rates ($k_i$) such that $k_x > k_y > k_z$, translating spin-populations into a triplet lifetime. Magnetic resonance, driven by a microwave field tuned to the energy splitting between two triplet sublevels, alters these spin state populations. This results in a change in the effective triplet lifetime, as for example shorter lived populations get shuffled into longer lived counterparts, this in turn alters the optical yield, enabling optical detected magnetic resonance (ODMR).

We use the setup to study both crystalline and disordered samples, using thermally evaporated thin-films and crystalline samples respectively. Figure \ref{fig1}c, d show the ODMR response of a crystal and film respectively when sweeping both field and microwave frequency about the $|T_x\rangle \leftrightarrow |T_z\rangle$ transition. We see in the crystalline sample the zero field transition splits into two peaks, whilst in the disordered thin-film the peak broadens into a powder spectra. The difference in these two samples highlights their structure, with the crystalline sample showing a high degree of order leading two only two distinct orientations of the pentacene molecules in the \textit{p}-terphenyl host and the film having a partially oriented structure with many different pentacene orientations\cite{lubert2018identifying}. Simulations of the ODMR considering the molecular ordering for both samples can be found in the supplemental information.

\section{Results}\label{sec2}
\section{Spatially resolved coherent control of a molecular thin-film}

\begin{figure*}[htb!]
\centerline{\includegraphics[
]{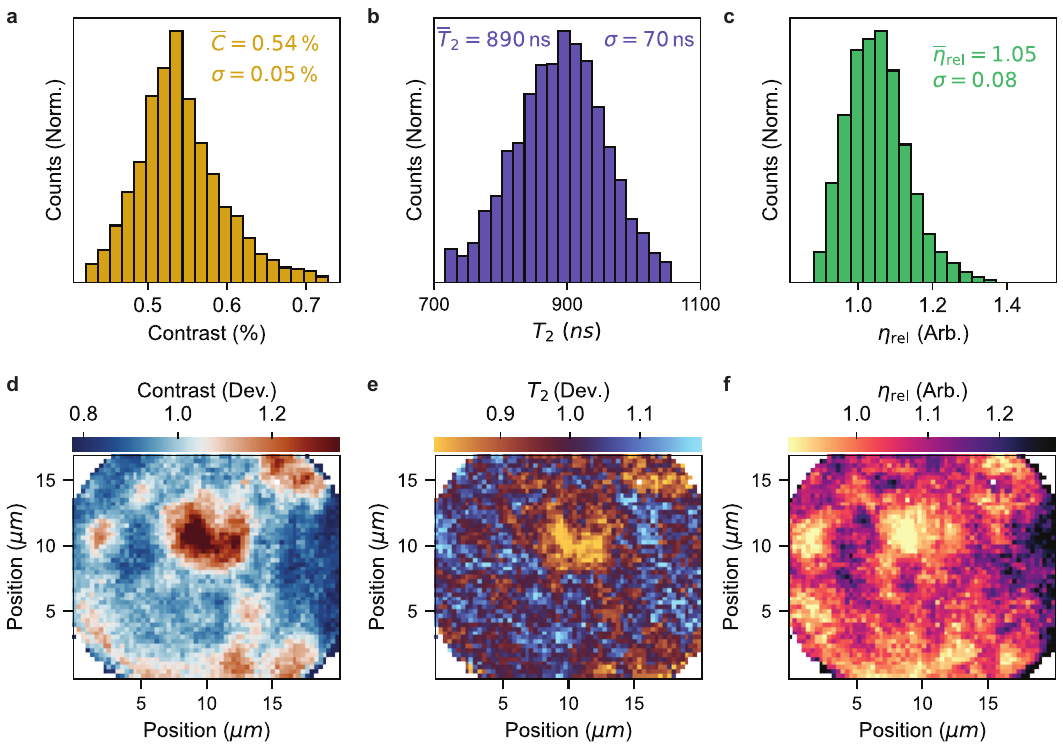}}
\caption{\textbf{Spatially resolved coherent control of a thin-film.} a) Histogram of the Hahn-echo measurement contrast across the film. b) Histogram of the coherence times, $T_2$, measured across the film using an optically detected Hahn-echo. c) Histogram of the sensitivity scale factors found by calculating the impact of contrast and coherence time variation in the film. d) Spatial map of the deviation in measurement contrast. e) Spatial map of the deviation in coherence time. f) Spatial map of the deviation in sensitivity scale factor.}\label{fig2}
\end{figure*}

Thin-films are an attractive approach to achieve high proximity quantum imaging, given that films can be reliably fabricated with nanometre precision \cite{kang2014enhancing} . This reduces the distance from the sensor to the target, allowing for the detection of smaller fields and an improvement in spatial resolution\cite{}. Additionally films are not limited to planar targets and evaporation can be performed onto non-planar substrates, allowing for 3 dimensional structures to be imaged. The trade-off here comes from the introduced disorder into the system, which will cause variation in the spin properties within the sensing layer and a non-uniform sensitivity\cite{mena2024inter, pappas2022resolving}. To understand the impact of disorder we apply a spatially resolved Hahn echo to a thermally evaporated thin-film of pentacene doped \textit{p}-terphenyl ($100\,$nm thickness, doped at 1\%).

Figure \ref{fig2}a shows a histogram of the measurement contrasts across the film, with an average contrast of $0.55$\,\% and a standard deviation of 0.05\,\%, leading to a variation of 9.8\,\%. The variation across the device is imaged in figure \ref{fig2}d, with the image showing the deviation in contrast ($C^{(i,j)}_{\mathrm{dev}} = C^{(i,j)}/C_{\mathrm{avg}}$) as a function of position. We note that the reported contrasts are significantly lower than previous reports for pentacene:\textit{p}-terpenyl\cite{mena2024room, singh2025room}. This is a consequence of our wide-field measurement technique rather than the film, we see a similar reduction in absolute contrast in the crystal samples below and in figure \ref{fig4} we see that when measured confocally the contrast is drastically improved.

Figure \ref{fig2}b shows the histogram for the coherence time, $T_2$, across our film. Here we find an average of $890$\,ns with a standard deviation of 70\,ns giving a variation of 7.9\,\%. The corresponding deviation map is shown in \ref{fig2}e, again clustering is observed here in the deviation across the device. Our average results agree with previously reported films, however the variation highlights the impact of disorder on the spin properties across our device, resulting in large variations of both contrast and coherence times. 

Given that for our excited state triplet the sensitivity is inversely proportional to both measurement contrast and coherence time\cite{barry2020sensitivity}, we can assess the impact of variation in contrast and coherence on the sensitivity of our film. We define a sensitivity adjustment factor, $\eta_{\mathrm{rel}}$, such that the sensitivity at a given point in the film is $\eta^{(i,j)} = \eta^{(i,j)}_{\mathrm{rel}} \cdot \eta_{\mathrm{avg}}$, where $\eta_{\mathrm{avg}}$ is the average sensitivity of the film and $\eta^{(i,j)}_{rel}$ is the local scale factor leading to an adjusted sensitivity, $\eta^{(i,j)}$, for a given point $(i,j)$ in the film (full calculation procedure can be found in the methods section). Figure \ref{fig2}c shows the histogram of scale factors ($\eta_{\mathrm{rel}}$) across our film, with variability of $7.6$\,\%.

\section{Spatially resolved coherent control of a molecular micro-crystal}

\begin{figure*}[htb!]
\centerline{\includegraphics[
]{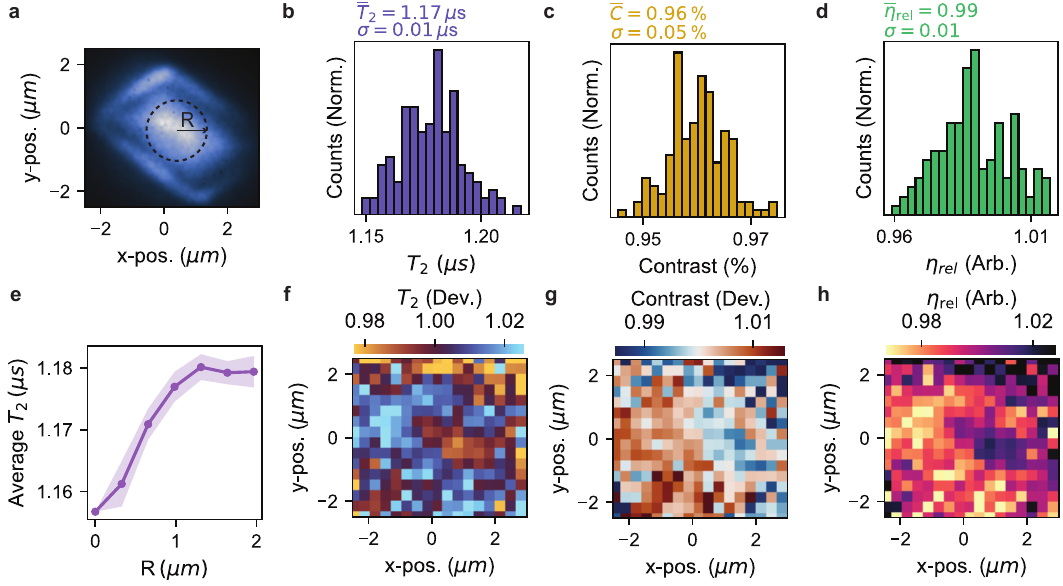}}
\caption{\textbf{Spatially resolved coherent control of a micro-crystal}. a) A photoluminescence image of the micro-crystal measured. b) Histogram of the coherence times measured within the same micro-crystal. c) Histogram of the measurement contrasts measured within the same micro-crystal. d) Histogram of the sensitivity scale factors measured within the same micro-crystal. e)Average coherence time as a function of radial distance ($R$) from the centre of the micro-crystal with errors represented by the shaded area. The radial distance is illustrated in figure a). f. Coherence time, $T_2$, mapped for the micro-crystal. The data has been plotted as deviation from the mean to highlight variation. g) Spatially resolved measurement contrast (C) mapped as deviations for the micro-crystal. h) Sensitivity scale factor, $\eta_\mathrm{rel}$, mapped as deviations across the micro-crystal.} \label{fig3}
\end{figure*}

Next we apply spatially resolved coherent control to an ordered system, using liquid phase exfoliation\cite{bailey2025revealing, bailey2025between}, we produce micro-flakes of pentacene:\textit{p}-terphenyl. These flakes are formed by sonicating a crystal with $0.1$\,\% pentacene concentration in acetone. The flakes are drop-cast onto a quartz slide and we isolate an individual flake to resolve its' spin properties. Figure \ref{fig3}a shows the photoluminescence of the flake used in the experiment.

Using the Hahn-echo sequence we are able to image both the coherence time and measurement contrast across the crystal. Figure \ref{fig3}f shows the resulting $T_2$ map and the corresponding histogram is in figure \ref{fig3}b. We find an average $T_2$ of $1.17$\,$\mu$s with a variation of $1$\,\%. The map shows an unexpected trend, with the edge of the crystal having a longer coherence time than the centre. To confirm this we average the measured coherence time with varying radial distances from the centre of the crystal (the radial distance ($R$) is indicated in figure \ref{fig3}b). Figure \ref{fig3}e shows the averaged $T_2$ plotted against the radial distance, with a clear trend that the edge of the crystal has a longer coherence time. 

The measurement contrasts and the corresponding deviation map can be found in figures \ref{fig3}c and g respectively. Here we find an average contrast of $0.96$\,\% with variation of $0.5$\,\%. Both the contrast and coherence time show less variation than in the film, suggesting that the disorder of the film is a key contributor to the variations in sensitivity and coherence. Our work also highlights the importance of considering the environment of the spin-bearing molecules when designing molecular quantum technologies\cite{wasielewski2026molecular}. Here we show that by changing the ordering of the same dopant-host system we start to see significantly varied spin properties across our devices. Therefore, finding suitable host molecules that can form well ordered crystalline environments could play a central role in unlocking a wider variety of spin-bearing molecules for quantum information science.

Finally, we use the contrast and the coherence time to spatially resolve the sensitivity in the crystalline micro-flake. The histogram of sensitivities in figure \ref{fig3}d shows an average sensitivity scale factor of $0.99$ with variations in sensitivity of $1.3$\%. As expected from the reduced variation in both contrast and coherence time, we find that the sensitivity is much more uniform across the crystalline flake. Mapping the sensitivity in figure \ref{fig3}h, we see the same structure appearing as in the coherence map with the edge being more sensitive than the centre.

\section{High-contrast optically detected coherent control of a nano-crystal}
\begin{figure}[htb!]
{\includegraphics[width=80mm]{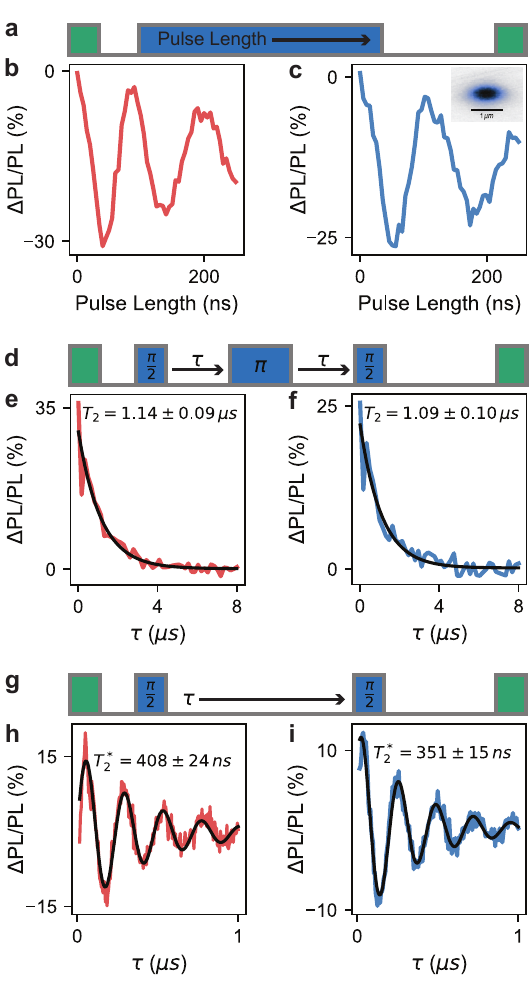}}
\caption{\textbf{Optically detected magnetic resonance of a bulk and nano-crystal.} a) Pulse sequence used for optically detected Rabi oscillations. b) Rabi oscillations measured on the bulk crystal with contrasts up to $30\,$\%. c) Rabi oscillations measured on the nanocrystal with contrasts up to $25\,$\%. d) Pulse sequence used for optically detected Hahn echo. e) Hahn echo measured on the bulk crystal with $1.14 \pm 0.09\,\mu$s coherence time and up to $35\,$\% contrast. c) Hahn echo measured on the nano-crystal with $1.09 \pm 0.10\,\mu$s coherence time and up to $25\,$\% contrast. g) Pulse sequence used for optically detected Ramsey fringes. h) Ramsey fringes measured on the bulk crystal with $T_2^* = 408 \pm 24\,$ns and up to $30\,$\% contrast. i) Ramsey fringes measured on the nano-crystal with $365 \pm 17\,$ns coherence time and up to $20\,$\% contrast. All measurements were performed on the $|T_x\rangle \leftrightarrow |T_z\rangle$ transition under ambient conditions.}\label{fig4}
\end{figure}

Motivated by the long coherence time and uniformity of the micro-crystal in figure \ref{fig3}, we next compare the optically detected coherent control of our bulk crystal to a single pentacene doped \textit{p}-terphenyl nanocrystal. This is done by locating an isolated nanocrystal and switching our detection scheme to replace the camera with a single photon counting avalanche photo-detector. We also remove the lens which previously acted to focus the excitation on the back focal plane of our objective, allowing us to collect light from a single focused point. An image of the nanocrystal is shown in the inset of figure \ref{fig4}c.

First we measure Rabi oscillations using the pulse sequence in figure \ref{fig4}a. The oscillations for the bulk crystal are shown in figure \ref{fig4}b and the corresponding measurement for the nanocrystal can be seen in figure \ref{fig4}c. We are able to achieve measurement contrasts of $30\%$ and $25\%$ in the bulk and nanocrystal respectively. These measurements exceed previous reports using pc:ptp nanocrystals by $20\%$\cite{ishiwata2025molecular} and show only a $5\%$ loss of contrast by reducing the crystal size below $1\,\mu m$.

By changing the pulse sequence to a Hahn-echo we are able to study the coherence of the systems. Using the sequence in figure \ref{fig4}d, we measure the decays shown in figures \ref{fig4}e and f for the bulk and nano-crystal respectively. Here we have the same measurement contrasts as in the Rabi oscillations and a $T_2$ of $1.14 \, \pm \, 0.09\,\mu$s for the bulk crystal and $1.09 \, \pm \, 0.1\,\mu$s for the nanocrystal.

Using a Ramsey sequence (with a detuning of $5\,$MHz from the XZ transition) as in figure \ref{fig4}g we are able to measure the spin dephasing, $T_2^*$ of the bulk and nano-crystals. The oscillation for the bulk crystal is shown in figure \ref{fig4}h with a contrast again of $30$\,\% and $T_2^* = 408\, \pm \,24$\,ns. Figure \ref{fig4}i displays the Ramsey oscillations for the nanocrystal, here we have $20\,$\% contrast and $T_2^* = 351 \, \pm \, 15$\,ns. Overall we find that even at quite high doping concentrations (0.1\,\% mol/mol) there is minimal impact on the contrast and coherence times of the molecular spins when reducing the nanocrystal size to sub-micrometre scales. This is promising from both a sensing stand-point and for implementing recently proposed levitated optomechanics experiments with molecular nano-crystals\cite{steiner2025optically}.

With the contrast and coherence times measured we can then estimate the loss in sensitivity incurred by reducing the crystal size. For AC sensing the relevant measurement is the Hahn-echo, where the nano-crystal dropped by a factor of 0.83 in contrast and 0.96 in coherence time, hence the sensitivity of the nanocrystal is impaired by a factor of 1.2 (remembering sensitivity is like golf). Likewise for DC sensing the Ramsey measurement had a contrast decrease of 0.67 and a dephasing time decrease of 0.89, leading to a sensitivity scaling of 1.7.

\section{Conclusion}\label{sec5}

Our work highlights the importance of understanding the effect of disorder within a quantum system on the coherence and contrast. Although the use of thin-films offers an approach to high-proximity sensing of both planar and non-planar structures, the introduced variability due to disorder will need to be accounted for when performing imaging using these materials. The increased uniformity of the micro-crystal, alongside the maintained coherence toward the edge of the flake offers an alternative to thin-films for high proximity sensing. Growing thin crystals\cite{wang2021vacuum,wu2022recent} on the surface of a sensing target could offer a higher performance alternative to thin-films. The micro- and nano-crystal results indicate that there is a much smaller impact on the coherence due to scaling down the crystal size than would be expected from solid state defects\cite{march2023long}. This suggests their future application as optically trapped sensors\cite{stewart2024optical} for imaging and in spin-based optomechanical experiments\cite{wood2021quantum}. We expect that finding novel methods to produce and place smaller doped molecular crystals to be a promising avenue for achieving higher resolution precision imaging.

\section{Experimental methods}

\subsection{Film preparation}

Thin films of p-terphenyl doped with pentacene were deposited onto quartz substrates via thermal co-evaporation of pentacene (Ossila, sublimated 99\%) and \textit{p}-terphenyl (TCI, >99\%) under vacuum in a commercial thermal evaporator (Kurt J. Lesker). To ensure the correct doping ratio we calibrated the evaporation rates of both pentacene and \textit{p}-terphenyl to ensure that the pentacene would evaporate at 0.1\% of the \textit{p}-terphenyl rate. The samples were then used unencapsulated and under ambient conditions.

\subsection{Crystal growth and exfoliation}

Crystals of p-terphenyl doped with pentacene were grown by solution recrystallization. Pentacene and p-terphenyl (0.1 mol \%) were dissolved in dichlorobenzene at 160$^\circ$C on a hotplate. The solution was then slowly cooled to room temperature over a period of 5 h, resulting in the formation of large crystals. The dichlorobenzene was subsequently decanted, and the crystals were washed with acetone to remove residual solvent before drying.

Micro-crystals were prepared by placing a single crystal in acetone and sonicating the suspension in an ultrasonic bath (VWR USC-THD Ultrasonic Cleaner) for 5 min. This process produced a stable suspension of micro-crystals in acetone. The micro-crystals were deposited onto quartz substrates by drop casting.

\subsection{Continuous wave optically detected magnetic resonance}

The sample is mounted between the poles of an electromagnet (Montana Instruments, Magneto-Optic Module) on a coplanar waveguide (CPW) which is used to generate microwave fields to drive resonance. The microwave signal is generated via a signal generator (Stanford Research Systems, SG396) and switched on and off at 211\,Hz using a microwave switch (Minicircuits, ZYSWA-2-50DR+), driven by an arbitrary waveform generator (AWG) (Swabian, Pulse Streamer). We use 520\,nm light from a solid-state laser (Hubner, Cobalt), which goes through a 550\,nm short pass filter (Thorlabs, FESH550) before being focussed on the sample via a achromatic doublet (Thorlabs, AC254-035-B). The photoluminescence is collected via the same lens and split from the laser path using a dichroic beam splitter (Thorlabs, DMLP550R) and focused onto a silicon photoreceiver (Femto, OE200Si) via a matching achromatic doublet. The signal from the photoreceiever is monitored via a lock-in amplifier (Stanford Research Systems, SR865A), allowing measurement of the lockin signal and the constant photoluminescence during the magnetic field sweep.

\subsection{Spatially resolved coherent control}

The sample is mounted on a CPW which is used to generate microwave fields for coherent control. The excitation, from a 520\,nm TTL controlled laser (Hubner, Cobolt 5051), is passed through a 550\,nm short pass filter (Thorlabs, FESH550) and a biconvex lens (Thorlabs, LB1437) which is one focal length away from the back-focal plane of the objective. The excitation then goes to a dichroic beam splitter (Thorlabs, DMLP550R) before going to a 0.9 NA objective (Ziess, EC Epiplan-Neofluar 100x). The photoluminescence is collected through the same objective, passes through the dichroic and through an additional longpass filter (Thorlabs, FELH550). This filtered light is then passed through a $200\,$mm (Thorlabs, LB4282) lens and imaged onto a sCMOS camera (Andor, iStar sCMOS). The microwave signal is generated via a signal generator with built in quadrature modulation for phase control (Stanford Research Systems, SG396), with pulse gating achieved via a microwave switch (Minicircuits, ZYSWA-2-50DR+) and amplification using a power amplifier (Minicircuits, HPA-272+). Timing control, used to pulse the laser and microwaves, as well as gate the camera is achieved using an AWG (Swabian, Pulse Streamer). Full details of the pulse sequences are outlined in our previous work\cite{mena2024room}.

\subsection{Optically detected coherent control}

For the nanocrystal and crystal measurement in figure \ref{fig4} the excitation and collection paths are altered. We remove the biconvex lens, meaning that collimated light is input into the objective for a diffraction limited excitation spot. On the collection side the tube lens is removed and we use a fibre coupled single photon counting avalanche photodiode (SNAPD) (Excelitas, SPCM- AQRH-14-FC), with the filtered photoluminescence being focused onto a single mode fibre and input into the SNAPD. We use a event counter (Swabian, Time Tagger Ultra) to count the pulses from the SNAPD. Full pulse sequences used are available in the supplementary information.

\subsection{Sensitivity Scale Factor and Data Analysis}

As we are in the measurement regime where the measurement overhead is much longer than the coherence time, $t_{\mathrm{overhead} } \gg T_2$, the sensitivity scales as $\eta^v_{\mathrm{AC}} \propto \sqrt{t_{\mathrm{overhead}}}/CT_2 \sqrt{n_{\mathrm{av}}c_s}$\cite{barry2020sensitivity}.  Here $C$ is the measurement contrast, $T_2$ is the coherence time, $n_{\mathrm{av}}$ is the doping ratio and $c_s$ is the collection efficiency. Our experiments let us directly probe the variations in contrast and coherence time, so we can then use these with the above equation to calculate the variation in sensitivity. We do this by first scaling both the contrast and the coherence time, scaling them as $C^{\mathrm{scale}}(i,j) = C(i,j)/C^{\mathrm{av}}$ and $T_2^{\mathrm{scale}}(i,j) = T_2(i,j)/T_2^\mathrm{av}$ respectively. Then the sensitivity scaling factor, $\eta_\mathrm{rel}^{(i,j)} = C^\mathrm{av} \cdot T_2^\mathrm{av}/C(i,j) \cdot T_2 (i,j)$, which is effectively a measure of how the sensitivity is influenced by the variation in both the measurement contrast and coherence time.  

We note that, for all analyses (including the contrast and coherence histograms), data points were removed based on the following criteria. First, points with poor fitting quality were excluded by applying a threshold of $r^2<0.85$. Second, outliers were removed by excluding values outside the 1st - 99th percentile range. Finally, for the microcrystal histograms in Fig. \ref{fig3}, data points lying outside the crystal were excluded using an intensity-based mask; a plot of this mask is provided in the Supplementary Materials.

\subsubsection{Fitting Functions}
Fitting is performed using the following functions:

Hahn-Echo: $y = Ce^{-2\tau/T_2} + B$

Ramsey: $y = Ce^{-\tau/T_2^*}cos(\omega\tau + \phi) + B$

\section*{Author contributions}

A.M, N.P.S and D.R.M contributed to the conceptualisation. A.M and N.P.S performed measurements and sample preparation. A.M and M.R.B analysed the data. A.M wrote the manuscript with inputs from all co-authors. Supervision, funding acquisition and resources were provided by D.R.M.

\section*{Acknowledgements}

A.M is the recipient of an Office of National Intelligence National
Intelligence Postdoctoral Grant (project number NIPG202510) funded by the
Australian Government. N.P.S acknowledges the support from an Australian Government Research Training Program (RTP) Scholarship. M.R.B acknowledges the support from the German Academic Exchange Service (DAAD) through the PROMOS scholarship.

\section*{Conflict of interest}

The authors declare no potential conflict of interests.

\bibliography{references}

@article{mena2024room,
  title={Room-temperature optically detected coherent control of molecular spins},
  author={Mena, Adrian and Mann, Sarah K and Cowley-Semple, Angus and Bryan, Emma and Heutz, Sandrine and McCamey, Dane R and Attwood, Max and Bayliss, Sam L},
  journal={Physical review letters},
  volume={133},
  number={12},
  pages={120801},
  year={2024},
  publisher={APS}
}

@article{mena2024inter,
  title={Inter-and Intra-Device Variation and Correlation of Hyperfine Interactions in Micron-Scale Organic Light-Emitting Diodes},
  author={Mena, A and Geng, R and Pappas, WJ and Maasoumi, F and McCamey, DR},
  journal={Advanced Sensor Research},
  volume={3},
  number={4},
  pages={2300087},
  year={2024},
  publisher={Wiley Online Library}
}

@article{pappas2022resolving,
  title={Resolving the Spatial Variation and Correlation of Hyperfine Spin Properties in Organic Light-Emitting Diodes},
  author={Pappas, William J and Geng, Rugang and Mena, Adrian and Baldacchino, Alexander J and Asadpoordarvish, Amir and McCamey, Dane R},
  journal={Advanced Materials},
  volume={34},
  number={11},
  pages={2104186},
  year={2022},
  publisher={Wiley Online Library}
}

@article{kopp2024luminescent,
  title={Luminescent organic triplet diradicals as optically addressable molecular qubits},
  author={Kopp, Sebastian M and Nakamura, Shunta and Phelan, Brian T and Poh, Yong Rui and Tyndall, Samuel B and Brown, Paige J and Huang, Yuheng and Yuen-Zhou, Joel and Krzyaniak, Matthew D and Wasielewski, Michael R},
  journal={Journal of the American Chemical Society},
  volume={146},
  number={40},
  pages={27935--27945},
  year={2024},
  publisher={American Chemical Society}
}

@article{singh2025room,
  title={Room-temperature quantum sensing with photoexcited triplet electrons in organic crystals},
  author={Singh, Harpreet and D'Souza, Noella and Zhong, Keyuan and Druga, Emanuel and Oshiro, Julianne and Blankenship, Brian and Montis, Riccardo and Reimer, Jeffrey A and Breeze, Jonathan D and Ajoy, Ashok},
  journal={Physical Review Research},
  volume={7},
  number={1},
  pages={013192},
  year={2025},
  publisher={American Physical Society}
}

@article{bayliss2020optically,
  title={Optically addressable molecular spins for quantum information processing},
  author={Bayliss, SL and Laorenza, DW and Mintun, PJ and Kovos, BD and Freedman, DE and Awschalom, DD},
  journal={Science},
  volume={370},
  number={6522},
  pages={1309--1312},
  year={2020},
  publisher={American Association for the Advancement of Science}
}

@article{bailey2025revealing,
  title={Revealing Localized Dark-Exciton Populations in 2D Perovskites via Magneto-Optical Microscopy},
  author={Bailey, Christopher G and Mena, Adrian and Leung, Tik Lun and Sloane, Nicholas P and Liao, Chwenhaw and McKenzie, David R and McCamey, Dane R and Ho-Baillie, Anita WY},
  journal={Advanced Energy Materials},
  pages={2501593},
  year={2025}
}

@article{tetienne2017quantum,
  title={Quantum imaging of current flow in graphene},
  author={Tetienne, Jean-Philippe and Dontschuk, Nikolai and Broadway, David A and Stacey, Alastair and Simpson, David A and Hollenberg, Lloyd CL},
  journal={Science advances},
  volume={3},
  number={4},
  pages={e1602429},
  year={2017},
  publisher={American Association for the Advancement of Science}
}

@article{scholten2022imaging,
  title={Imaging current paths in silicon photovoltaic devices with a quantum diamond microscope},
  author={Scholten, SC and Abrahams, GJ and Johnson, BC and Healey, AJ and Robertson, IO and Simpson, DA and Stacey, A and Onoda, S and Ohshima, T and Kho, TC and others},
  journal={Physical Review Applied},
  volume={18},
  number={1},
  pages={014041},
  year={2022},
  publisher={APS}
}

@article{gao2024nanotube,
  title={Nanotube spin defects for omnidirectional magnetic field sensing},
  author={Gao, Xingyu and Vaidya, Sumukh and Dikshit, Saakshi and Ju, Peng and Shen, Kunhong and Jin, Yuanbin and Zhang, Shixiong and Li, Tongcang},
  journal={Nature Communications},
  volume={15},
  number={1},
  pages={7697},
  year={2024},
  publisher={Nature Publishing Group UK London}
}

@article{dolde2011electric,
  title={Electric-field sensing using single diamond spins},
  author={Dolde, Florian and Fedder, Helmut and Doherty, Marcus W and N{\"o}bauer, Tobias and Rempp, Florian and Balasubramanian, Gopalakrishnan and Wolf, Thomas and Reinhard, Friedemann and Hollenberg, Lloyd CL and Jelezko, Fedor and others},
  journal={Nature Physics},
  volume={7},
  number={6},
  pages={459--463},
  year={2011},
  publisher={Nature Publishing Group UK London}
}

@article{barson2021nanoscale,
  title={Nanoscale vector electric field imaging using a single electron spin},
  author={Barson, Michael SJ and Oberg, Lachlan M and McGuinness, Liam P and Denisenko, Andrej and Manson, Neil B and Wrachtrup, Jorg and Doherty, Marcus W},
  journal={Nano Letters},
  volume={21},
  number={7},
  pages={2962--2967},
  year={2021},
  publisher={ACS Publications}
}

@article{gottscholl2020initialization,
  title={Initialization and read-out of intrinsic spin defects in a van der Waals crystal at room temperature},
  author={Gottscholl, Andreas and Kianinia, Mehran and Soltamov, Victor and Orlinskii, Sergei and Mamin, Georgy and Bradac, Carlo and Kasper, Christian and Krambrock, Klaus and Sperlich, Andreas and Toth, Milos and others},
  journal={Nature materials},
  volume={19},
  number={5},
  pages={540--545},
  year={2020},
  publisher={Nature Publishing Group UK London}
}

@article{stern2022room,
  title={Room-temperature optically detected magnetic resonance of single defects in hexagonal boron nitride},
  author={Stern, Hannah L and Gu, Qiushi and Jarman, John and Eizagirre Barker, Simone and Mendelson, Noah and Chugh, Dipankar and Schott, Sam and Tan, Hoe H and Sirringhaus, Henning and Aharonovich, Igor and others},
  journal={Nature communications},
  volume={13},
  number={1},
  pages={618},
  year={2022},
  publisher={Nature Publishing Group UK London}
}

@article{wolf2015subpicotesla,
  title={Subpicotesla diamond magnetometry},
  author={Wolf, Thomas and Neumann, Philipp and Nakamura, Kazuo and Sumiya, Hitoshi and Ohshima, Takeshi and Isoya, Junichi and Wrachtrup, J{\"o}rg},
  journal={Physical Review X},
  volume={5},
  number={4},
  pages={041001},
  year={2015},
  publisher={APS}
}

@article{stewart2024optical,
  title={Optical tweezers assembled nanodiamond quantum sensors},
  author={Stewart, Adam and Zhu, Ying and Liu, Yiting and Simpson, David A and Reece, Peter J},
  journal={Nano Letters},
  volume={24},
  number={39},
  pages={12188--12195},
  year={2024},
  publisher={ACS Publications}
}

@article{wood2021quantum,
  title={Quantum control of nuclear-spin qubits in a rapidly rotating diamond},
  author={Wood, Alexander A and Goldblatt, Russell M and Scholten, Robert E and Martin, Andy M},
  journal={Physical Review Research},
  volume={3},
  number={4},
  pages={043174},
  year={2021},
  publisher={APS}
}

@article{march2023long,
  title={Long spin coherence and relaxation times in nanodiamonds milled from polycrystalline 12 C diamond},
  author={March, James E and Wood, Benjamin D and Stephen, Colin J and Fervenza, Laura Dur{\'a}n and Breeze, Ben G and Mandal, Soumen and Edmonds, Andrew M and Twitchen, Daniel J and Markham, Matthew L and Williams, Oliver A and others},
  journal={Physical Review Applied},
  volume={20},
  number={4},
  pages={044045},
  year={2023},
  publisher={APS}
}

@article{wang2021vacuum,
  title={Vacuum processed large area doped thin-film crystals: A new approach for high-performance organic electronics},
  author={Wang, S-J and Sawatzki, M and Kleemann, H and Lashkov, I and Wolf, D and Lubk, A and Talnack, F and Mannsfeld, S and Krupskaya, Y and B{\"u}chner, B and others},
  journal={Materials Today Physics},
  volume={17},
  pages={100352},
  year={2021},
  publisher={Elsevier}
}

@article{wu2022recent,
  title={Recent developments of nanodiamond quantum sensors for biological applications},
  author={Wu, Yingke and Weil, Tanja},
  journal={Advanced Science},
  volume={9},
  number={19},
  pages={2200059},
  year={2022},
  publisher={Wiley Online Library}
}

@article{singh2025high,
  title={High sensitivity pressure and temperature quantum sensing in pentacene-doped p-terphenyl single crystals},
  author={Singh, Harpreet and D’Souza, Noella and Garrett, Joseph and Singh, Angad and Blankenship, Brian and Druga, Emanuel and Montis, Riccardo and Tan, Liang Z and Ajoy, Ashok},
  journal={Nature Communications},
  volume={16},
  number={1},
  pages={10530},
  year={2025},
  publisher={Nature Publishing Group UK London}
}

@article{feder2025fluorescent,
  title={A fluorescent-protein spin qubit},
  author={Feder, Jacob S and Soloway, Benjamin S and Verma, Shreya and Geng, Zhi Z and Wang, Shihao and Kifle, Bethel B and Riendeau, Emmeline G and Tsaturyan, Yeghishe and Weiss, Leah R and Xie, Mouzhe and others},
  journal={Nature},
  volume={645},
  number={8079},
  pages={73--79},
  year={2025},
  publisher={Nature Publishing Group UK London}
}

@article{gorgon2023reversible,
  title={Reversible spin-optical interface in luminescent organic radicals},
  author={Gorgon, Sebastian and Lv, Kuo and Gr{\"u}ne, Jeannine and Drummond, Bluebell H and Myers, William K and Londi, Giacomo and Ricci, Gaetano and Valverde, Danillo and Tonnel{\'e}, Claire and Murto, Petri and others},
  journal={Nature},
  volume={620},
  number={7974},
  pages={538--544},
  year={2023},
  publisher={Nature Publishing Group UK London}
}

@article{chowdhury2025room,
  title={Room temperature optical control of spin states in organic diradicals},
  author={Chowdhury, Rituparno and Inglis, Alistair and Walker, Lucy E and Murto, Petri and Delpiano-Cordeiro, Chiara and Morrison, Colin and Panjwani, Naitik A and Fu, Yao and Sun, Yan and Zhou, Wei and others},
  journal={arXiv preprint arXiv:2510.09440},
  year={2025}
}

@article{ishiwata2025molecular,
  title={Molecular quantum nanosensors functioning in living cells},
  author={Ishiwata, Hitoshi and Song, Jiarui and Shigeno, Yoko and Nishimura, Koki and Yanai, Nobuhiro},
  year={2025}
}

@article{mann2025chemically,
  title={Chemically Tuning Room Temperature Pulsed Optically Detected Magnetic Resonance},
  author={Mann, Sarah K and Cowley-Semple, Angus and Bryan, Emma and Huang, Ziqiu and Heutz, Sandrine and Attwood, Max and Bayliss, Sam L},
  journal={Journal of the American Chemical Society},
  year={2025},
  publisher={ACS Publications}
}

@article{kopp2025optically,
  title={Optically Detected Coherent Spin Control of Organic Molecular Color Center Qubits},
  author={Kopp, Sebastian M and Nakamura, Shunta and Poh, Yong Rui and Peinkofer, Kathryn R and Phelan, Brian T and Yuen-Zhou, Joel and Krzyaniak, Matthew D and Wasielewski, Michael R},
  journal={Journal of the American Chemical Society},
  year={2025},
  publisher={ACS Publications}
}

@article{abrahams2024quantum,
  title={Quantum spin resonance in engineered magneto-sensitive fluorescent proteins enables multi-modal sensing in living cells},
  author={Abrahams, Gabriel and {\v{S}}tuhec, Ana and Spreng, Vincent and Henry, Robin and Kempf, Idris and James, Jessica and Sechkar, Kirill and Stacey, Scott and Trelles-Fernandez, Vicente and Antill, Lewis M and others},
  journal={bioRxiv},
  pages={2024--11},
  year={2024},
  publisher={Cold Spring Harbor Laboratory}
}

@article{sloop1981electron,
  title={Electron spin echoes of a photoexcited triplet: Pentacene in p-terphenyl crystals},
  author={Sloop, David J and Yu, Hsiang-Lin and Lin, Tien-Sung and Weissman, SI},
  journal={The Journal of Chemical Physics},
  volume={75},
  number={8},
  pages={3746--3757},
  year={1981},
  publisher={American Institute of Physics}
}

@article{li2025robust,
  title={Robust AC vector sensing at zero magnetic field with pentacene},
  author={Li, Boning and Heller, Garrett and Yong, Jungbae and Ungar, Alexander and Tang, Hao and Wang, Guoqing and Hautle, Patrick and Quan, Yifan and Cappellaro, Paola},
  journal={arXiv preprint arXiv:2512.06272},
  year={2025}
}

@article{lubert2018identifying,
  title={Identifying triplet pathways in dilute pentacene films},
  author={Lubert-Perquel, Daphn{\'e} and Salvadori, Enrico and Dyson, Matthew and Stavrinou, Paul N and Montis, Riccardo and Nagashima, Hiroki and Kobori, Yasuhiro and Heutz, Sandrine and Kay, Christopher WM},
  journal={Nature communications},
  volume={9},
  number={1},
  pages={4222},
  year={2018},
  publisher={Nature Publishing Group UK London}
}

@article{kang2014enhancing,
  title={Enhancing 2D growth of organic semiconductor thin films with macroporous structures via a small-molecule heterointerface},
  author={Kang, Boseok and Jang, Moonjeong and Chung, Yoonyoung and Kim, Haena and Kwak, Sang Kyu and Oh, Joon Hak and Cho, Kilwon},
  journal={Nature communications},
  volume={5},
  number={1},
  pages={4752},
  year={2014},
  publisher={Nature Publishing Group UK London}
}

@article{kucsko2013nanometre,
  title={Nanometre-scale thermometry in a living cell},
  author={Kucsko, Georg and Maurer, Peter C and Yao, Norman Ying and Kubo, MICHAEL and Noh, Hyun Jong and Lo, Po Kam and Park, Hongkun and Lukin, Mikhail D},
  journal={Nature},
  volume={500},
  number={7460},
  pages={54--58},
  year={2013},
  publisher={Nature Publishing Group UK London}
}

@article{fujiwara2020real,
  title={Real-time nanodiamond thermometry probing in vivo thermogenic responses},
  author={Fujiwara, Masazumi and Sun, Simo and Dohms, Alexander and Nishimura, Yushi and Suto, Ken and Takezawa, Yuka and Oshimi, Keisuke and Zhao, Li and Sadzak, Nikola and Umehara, Yumi and others},
  journal={Science advances},
  volume={6},
  number={37},
  pages={eaba9636},
  year={2020},
  publisher={American Association for the Advancement of Science}
}

@article{lovchinsky2016nuclear,
  title={Nuclear magnetic resonance detection and spectroscopy of single proteins using quantum logic},
  author={Lovchinsky, Igor and Sushkov, AO and Urbach, Elana and de Leon, Nathalie P and Choi, Soonwon and De Greve, Kristiaan and Evans, Ruffin and Gertner, Rona and Bersin, Eric and M{\"u}ller, C and others},
  journal={Science},
  volume={351},
  number={6275},
  pages={836--841},
  year={2016},
  publisher={American Association for the Advancement of Science}
}

@article{barry2020sensitivity,
  title={Sensitivity optimization for NV-diamond magnetometry},
  author={Barry, John F and Schloss, Jennifer M and Bauch, Erik and Turner, Matthew J and Hart, Connor A and Pham, Linh M and Walsworth, Ronald L},
  journal={Reviews of Modern Physics},
  volume={92},
  number={1},
  pages={015004},
  year={2020},
  publisher={APS}
}

@article{bertran2020light,
  title={Light-induced triplet--triplet electron resonance spectroscopy},
  author={Bertran, Arnau and Henbest, Kevin B and De Zotti, Marta and Gobbo, Marina and Timmel, Christiane R and Di Valentin, Marilena and Bowen, Alice M},
  journal={The journal of physical chemistry letters},
  volume={12},
  number={1},
  pages={80--85},
  year={2020},
  publisher={ACS Publications}
}

@article{doherty2014electronic,
  title={Electronic properties and metrology applications of the diamond NV- center under pressure},
  author={Doherty, Marcus W and Struzhkin, Viktor V and Simpson, David A and McGuinness, Liam P and Meng, Yufei and Stacey, Alastair and Karle, Timothy J and Hemley, Russell J and Manson, Neil B and Hollenberg, Lloyd CL and others},
  journal={Physical review letters},
  volume={112},
  number={4},
  pages={047601},
  year={2014},
  publisher={APS}
}

@article{wood20243d,
  title={3D-mapping and manipulation of photocurrent in an optoelectronic diamond device},
  author={Wood, Alexander A and McCloskey, Daniel J and Dontschuk, Nikolai and Lozovoi, Artur and Goldblatt, Russell M and Delord, Tom and Broadway, David A and Tetienne, Jean-Philippe and Johnson, Brett C and Mitchell, Kaih T and others},
  journal={Advanced Materials},
  volume={36},
  number={40},
  pages={2405338},
  year={2024},
  publisher={Wiley Online Library}
}

@article{steiner2025optically,
  title={Optically Hyperpolarized Materials for Levitated Optomechanics},
  author={Steiner, Marit OE and Pedernales, Julen S and Plenio, Martin B},
  journal={Quantum},
  volume={9},
  pages={1928},
  year={2025},
  publisher={Verein zur F{\"o}rderung des Open Access Publizierens in den Quantenwissenschaften}
}

@article{wu2025fluorescent,
  title={Fluorescent Nanodiamonds Based Theranostic Platform for pH-Sensitive Drug Delivery and Quantum Sensing},
  author={Wu, Kaiqui and Fan, Siyu and Zhang, Yue and Woudstra, Willem and Mulder, Thomas and Boscher Navarro, Laurens and van Dijken, Jur and Loos, Katja and Schirhagl, Romana},
  journal={Advanced Functional Materials},
  pages={e14294},
  year={2025},
  publisher={Wiley Online Library}
}

@article{zhang2024dynamics,
  title={Dynamics for high-sensitivity detection of free radicals in primary bronchial epithelial cells upon stimulation with cigarette smoke extract},
  author={Zhang, Y and Sigaeva, A and Fan, S and Norouzi, N and Zheng, X and Heijink, IH and Slebos, DJ and Pouwels, SD and Schirhagl, R},
  journal={Nano Letters},
  volume={24},
  number={31},
  pages={9650--9657},
  year={2024},
  publisher={ACS Publications}
}

@article{doherty2013nitrogen,
  title={The nitrogen-vacancy colour centre in diamond},
  author={Doherty, Marcus W and Manson, Neil B and Delaney, Paul and Jelezko, Fedor and Wrachtrup, J{\"o}rg and Hollenberg, Lloyd CL},
  journal={Physics Reports},
  volume={528},
  number={1},
  pages={1--45},
  year={2013},
  publisher={Elsevier}
}

@article{christle2015isolated,
  title={Isolated electron spins in silicon carbide with millisecond coherence times},
  author={Christle, David J and Falk, Abram L and Andrich, Paolo and Klimov, Paul V and Hassan, Jawad Ul and Son, Nguyen T and Janz{\'e}n, Erik and Ohshima, Takeshi and Awschalom, David D},
  journal={Nature materials},
  volume={14},
  number={2},
  pages={160--163},
  year={2015},
  publisher={Nature Publishing Group UK London}
}

@article{miao2019electrically,
  title={Electrically driven optical interferometry with spins in silicon carbide},
  author={Miao, Kevin C and Bourassa, Alexandre and Anderson, Christopher P and Whiteley, Samuel J and Crook, Alexander L and Bayliss, Sam L and Wolfowicz, Gary and Thiering, Gerg{\H{o}} and Udvarhelyi, P{\'e}ter and Iv{\'a}dy, Viktor and others},
  journal={Science Advances},
  volume={5},
  number={11},
  pages={eaay0527},
  year={2019},
  publisher={American Association for the Advancement of Science}
}

@article{bailey2025between,
  title={Between the Nanosheets: Enhancing Electron--Hole Exchange Interaction for Room-Temperature Magneto-Photoluminescence in Liquid-phase-exfoliated 2D Perovskite},
  author={Bailey, Christopher G and Sloane, Nicholas P and Leung, Tik Lun and Liao, Chwenhaw and Mena, Adrian and de Clercq, Damon M and Yi, Jianpeng and Palomba, Stefano and Nielsen, Michael P and McKenzie, David R and others},
  journal={ACS nano},
  year={2025},
  publisher={American Chemical Society}
}

@article{wasielewski2026molecular,
  title={Molecular Diradical Spin Qubits in a Crystalline Host as a Platform for Quantum Sensing},
  author={Wasielewski, Michael and Kopp, Sebastian and Palmer, Jonathan and Phelan, Brian and Peinkofer, Kathryn and Nakamura, Shunta and Krzyaniak, Matthew and others},
  year={2026}
}


\end{document}